# Upgrades to the Event Simulation and Reconstruction for the Fermi Large Area Telescope

Leon Rochester, Tracy Usher
*SLAC, Stanford, CA 94025, USA*

Robert P. Johnson, Bill Atwood
*SCIPP, UC Santa Cruz, Santa Cruz, CA 95064*

on behalf of the Fermi Large Area Telescope Collaboration

The pre-launch event simulation and reconstruction performed beyond our expectations for real data, essentially without modification, and made possible the immediate start of science analysis. But the on-orbit data exhibit unanticipated features that necessitate upgrades to both the simulation (essentially complete) and the reconstruction (ongoing). The major new effect encountered on orbit is the presence of "ghosts," that is, remnant detector response to particles passing through the detector before the particle that triggered the event. These ghosts appear primarily in the form of extra tracks and/or energy deposits. As part of this upgrade, we plan to enhance our ability to discriminate against background particles by introducing additional analysis during the reconstruction phase. We present a description of the effect of ghosts, and of the work needed to deal with them, done and planned, as well as some other ideas for improving the reconstruction.

## 1. THE LARGE AREA TELESCOPE

The Large Area Telescope (LAT) is a pair-conversion spectrometer based on the conversion of gamma-rays into electron-positron pairs. It is arranged in a 4×4 array of 16 identical towers. Each tower consists of tracker stack of silicon micro-strip detectors and tungsten conversion foils (TKR), above a segmented CsI calorimeter (CAL). Together they serve to reconstruct the gamma-ray direction and energy. A custom-designed data-acquisition module is located below the calorimeter. The tracker towers are covered by a segmented anti-coincidence system (ACD), consisting of panels of plastic scintillator read out by wave-shifting fibers and photo-multiplier tubes. The ACD provides our primary rejection of cosmic-ray charged-particle background (electrons, protons, and heavier nuclei).

The ACD is surrounded by a thermal blanket and micrometeor shield. An aluminum grid supports the detector modules, the data-acquisition system and the computers, which are all located below the CAL modules. A more detailed description is found in [1].

## 2. OUR ENCOUNTER WITH GHOSTS

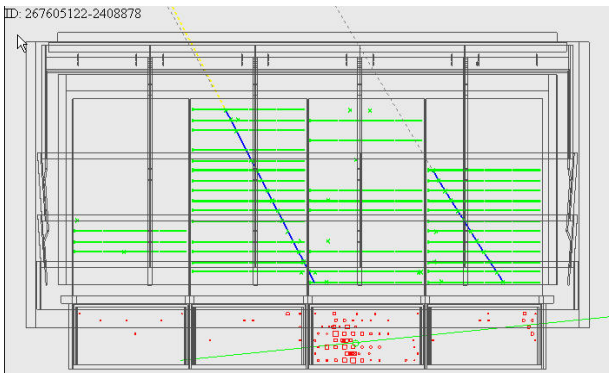

Figure 1: An unexpected event.

Once the LAT was on orbit, we started processing data using reconstruction code developed over the years before the launch. This code, described in [2], performed very well, and allowed us to begin science analysis immediately.

But in the data stream coming down from the LAT, we encountered, at the level of a few percent, events that were of a type that we hadn't anticipated. These events contained two or more tracks, clearly separated from each other. Fig. 1 shows an example of such an event.

The green horizontal line segments and Xs show the hit "x" and "y" strips on the silicon detectors in the tracker (TKR), and the blue lines show the found tracks. The "best" track is indicated by the dotted yellow line projecting upward from the head of the track. The red squares at the bottom indicate energy deposited in the calorimeter (CAL); the green box shows the centroid of the deposited energy, and the nearly horizontal green line shows the inferred axis of the shower.

In its present form, this event is essentially useless. First, we don't know what to make of the two tracks; could this be a shower of some sort? Second, the current energy calculation lumps all the deposited energy into a single cluster, whose direction is obviously incorrect in this case, as is, to a lesser extent, its centroid.

The origin of these extra tracks turns out to be the finite response time for each of the subsystems in the LAT. For example, the signal in the TKR of a hit from a cosmic ray track will be present for at least ten microseconds after it passes through the detector, and longer for higher-Z particles. Such latent hits are nicely captured by the periodic trigger (PT) in the LAT. This trigger is generated internally at a uniform rate of two per second, completely independent of the actual state of the detector, and thus gives us a picture of the background environment on orbit.

The extra signals are tracks that crossed the detector at some time before the trigger, and whose detector response has decayed by the time of the trigger. This is borne out by the timing information recorded with the event. We call these tracks "ghosts."

Most of the PT events are essentially empty. But Fig. 2 shows a sample of PT events with significant ghosts.

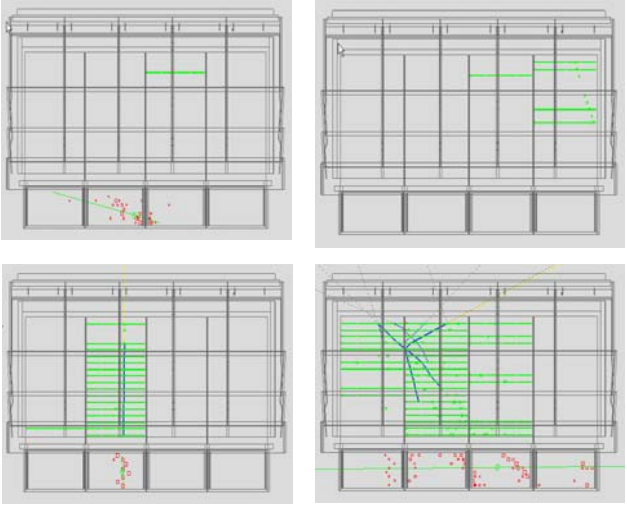

Figure 2: A sample of ghost events

The first has a signal in the CAL, but none in the TKR; the second, vice versa; and the bottom two have signals in both the CAL and TKR, as well as reconstructed tracks. In principle, there can also be remnant signals in the anti-coincidence detector (ACD), but generally these have a shorter time constant and decay quickly; hence, they are not prominently seen in these events.

The charged-particle background rate ($R_B$) in the detector depends mainly on the local magnetic field. Fig. 3 shows $R_B$ as a function of earth coordinates.

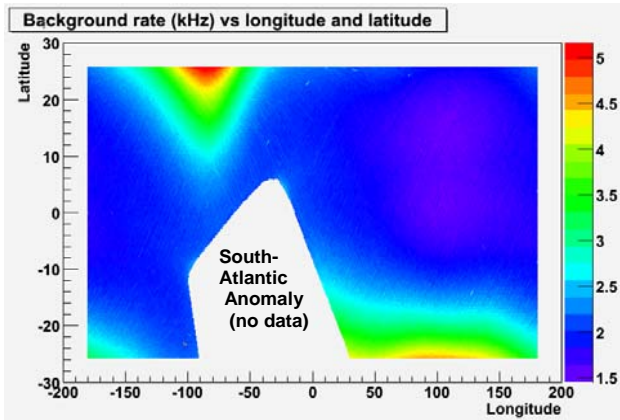

Figure 3: $R_B$ vs. earth coordinates.

Using the PTs, we can measure the potential contamination of real events due to ghosts as a function of the measured $R_B$. Fig. 4 shows the results of this measurement.

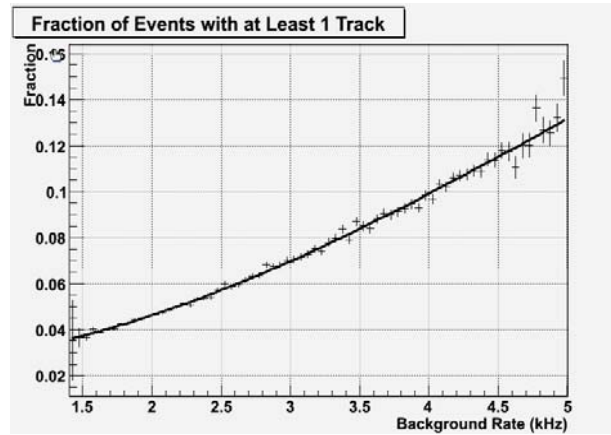

Figure 4: Fraction of PT events with at least one found track as a function of $R_B$

## 3. UPGRADING THE SIMULATION

Because a ghost can corrupt a gamma event so that the current analysis no longer recognizes it, the presence of ghosts affects the efficiency for detection of gammas, in a way that depends on the charged-particle trigger rate and the energy of the triggering event. We hope to ultimately recover many of these events. But as a first step, we decided to try simply to account for these ghosts in our Monte Carlo (MC) simulation. To this end, we modified the simulation so that we could overlay a real PT event on top of each simulated gamma event. The resulting events should be very similar to our actual gamma events.

This indeed turned out to be the case. In Fig. 5, we show the effective area of these "haunted" MC gamma-rays relative to that for unmodified ones, as a function of the energy of the gamma. As expected, the presence of the ghosts tends to decrease the effective area, and more so for low-energy gammas.

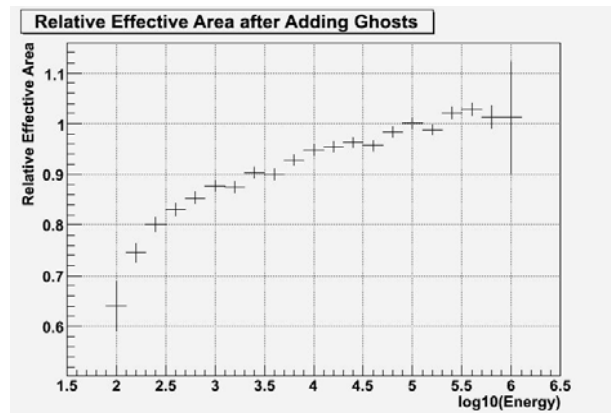

Figure 5: Relative effective area, integrated over all angles, for MC events with superimposed ghosts vs. the energy of the incident gamma.

As a check of the simulation, P. Bruel performed independent measurements of the gamma-ray flux of the Vela pulsar using data taken from separate ranges of $R_B$. The results are displayed in Fig. 6.

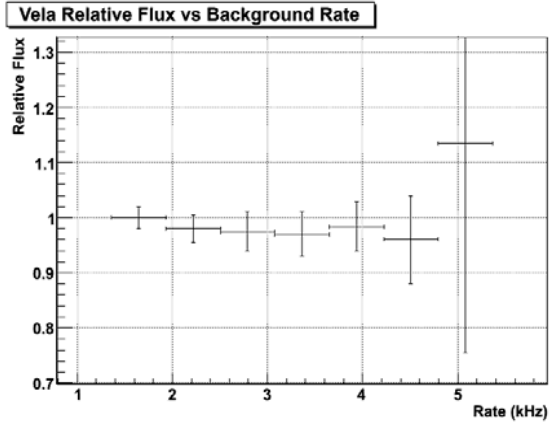

Figure 6: Measured flux of the Vela pulsar vs. $R_B$, based on six months of *Fermi* data

Each point in the plot shows the result of a flux calculation using the effective area for that $R_B$ bin, as determined by the simulation. The measured fluxes are found not to depend on the background rate of the sample. Similar results are obtained for the other parameters of the fit, for example, the power-law index.

## 4. FINDING CAL CLUSTERS

The first step, essential for developing a more effective event reconstruction, is to identify isolated clusters of energy deposition in the CAL. This seems straightforward, but turns out to be challenging.

One complication comes from the substantial gaps between the CAL modules in adjacent towers. We have developed a strategy to link up the sections of a cluster that cross towers.

Another issue is the treatment of single isolated crystals. For now we count each of these as a cluster, but we may eventually learn how to attach them to the appropriate nearby extended cluster.

Finally, we will need to re-tune the reconstruction of the energy associated with each event, since removing isolated crystals from the overall cluster will involve an apparent decrease of measured energy.

The displays below show an example of clustering. In Fig. 7, the calorimeter response is analyzed with the current single-cluster algorithm.

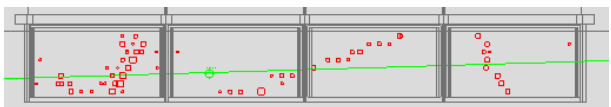

Figure 7: Clustering of a complex event using the current single-cluster algorithm. The green line shows the axis of the resulting "cluster."

Fig. 8 shows the same event with the newly developed multi-cluster algorithm. The resulting cluster axes are color-coded roughly in order of decreasing energy: red, orange, yellow, green, blue and violet. As indicated before, some of the single-crystal "clusters" are most likely associated with a nearby multi-crystal cluster.

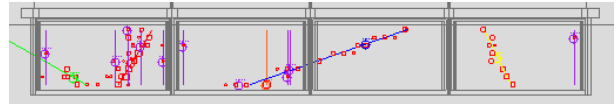

Figure 8: The event of Fig. 7 analyzed with the multi-cluster algorithm. Note the successful association of crystals across a tower boundary.

This first attempt has yielded promising results, but we will no doubt have to refine it as we better understand the constraints of the problem.

## 5. TAGGING GHOST TRACKS WITH TRIGGER INFO

It turns out that many of the TKR hits produced by ghost tracks can be easily identified. This is because they are out of time and fail to trigger at the level of individual TKR planes or full towers. The triggering efficiency of normal tracker hits is well above 99%; thus a track whose component hits should have generated a trigger, but didn't, can be confidently tagged as a ghost.

Fig. 9 shows an event with a normal track and one that consists of ghost hits tagged using trigger information.

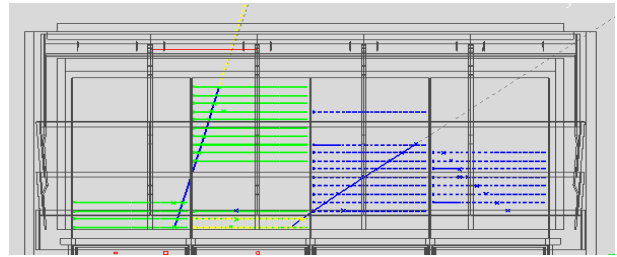

Figure 9: Event tagged with trigger information

The tagged ghost hits are colored blue. The hits colored yellow are ones that happen to produce plane triggers (and thus are not tagged), but which end up on the tagged track. Such hits are very likely to come from the ghost particle.

Also note a third track on the right, which was not found by our current code. We hope to improve on this situation in the next round!

## 6. USING THE NEW INFORMATION

In Fig. 10, we see the same event shown in Fig. 1, but with the added information coming from the upgrades to the TKR and CAL analyses:

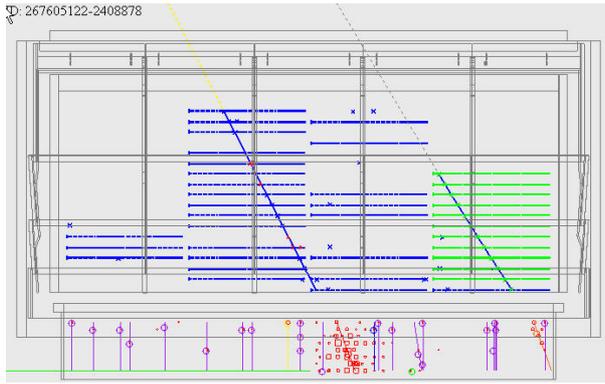

Figure 10: Event of Fig. 1 reanalyzed with upgraded algorithms

We see that the track on the left, which was originally chosen as the best track, is in fact a ghost, and that the most energetic CAL cluster in the event (1726 MeV) is associated with this ghost track. The cluster associated with the triggering track, on the right, has an energy of 129 MeV. Note that the axes of both of the CAL clusters line up with their corresponding TKR tracks.

Judging from the TKR time-over-threshold signals and the energy deposit in the CAL, the triggering track appears to be an upward-going proton that stops in the middle of the tracker. The ghost appears to be the remnant of a heavy ion.

Although unraveling this particular event was relatively straightforward, the typical haunted event will be more complicated.

## 7. PUTTING IT ALL TOGETHER

Solving the remaining piece of the puzzle will require the development of strategies to associate tracks and clusters and to characterize the particles in complex events. This will also include accounting for the events that remain too complicated to resolve.

To this end, we will have to redesign the current reconstruction algorithm to be more flexible, and allow for iteration and multiple paths.

While the specific implementation of our reconstruction algorithm doesn't yet have these features, the framework itself is designed to allow this flexibility.

## 8. OTHER UPGRADES

The original aim of our reconstruction strategy was to find gammas, not background. But as we understand more about our actual problem, we've come to understand that we must be able to do both well.

### 8.1. Identifying more background tracks

Since the hit efficiency of the TKR is so high, it's very unlikely that more than one or two hits will be missing on a normal track. So our track-finding is tuned to reject tracks with multiple gaps. But in many cases, the hits on a ghost track have decayed to the point where a substantial number of them are missing, even though it's obvious by eye that the remaining hits belong to a track. The upper-right event in the PT sampler (Fig. 2) contains such a track. For the reason cited above, these defective tracks are usually not found. The hits themselves are often tagged by their failure to trigger, but recognizing them as tracks may help us to eliminate any CAL cluster that they point to.

### 8.2. ACD-seeded track-finding

As noted earlier, we've seen events with seemingly findable TKR tracks that are not actually found. An important class of such tracks consists of those that point to a struck ACD tile. These are missed background tracks, which may have actually triggered the event.

Fig. 11 shows one such missed track. Since such particles can be a source of background, we need to modify our software to better identify and deal with them.

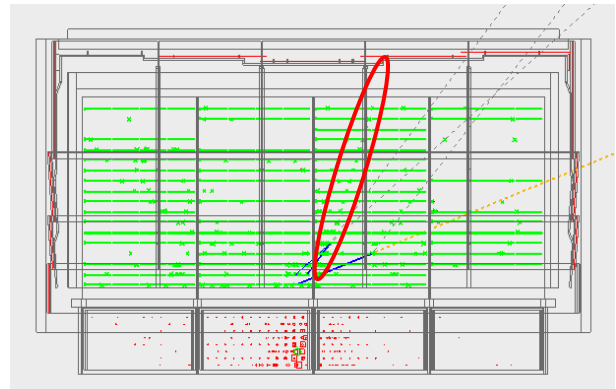

Figure 11: Event showing a track pointing to a struck tile, one not found by our current track-finding scheme

The existing track-finding looks for tracks along rays centered on the direction and position of CAL clusters. We intend to add a search method that will work inward from TKR hits near the sides and top of the LAT, close to hit ACD tiles.

### 8.3. Linked-vector and tree-based track-finding

Our current track-finding is based on local track following and tends to find many track stubs in the middle of electromagnetic (EM) showers. Only the one or two tracks at the head of the shower give us precise information about the direction of the gamma.

The remaining tracks are essentially useless, except possibly to indicate the presence of a shower.

We are exploring a complementary approach, in which we try to link up local vectors formed from x-y pairs of hits in vertically adjacent silicon detectors into a global structure.

Two examples are shown in Fig. 12.

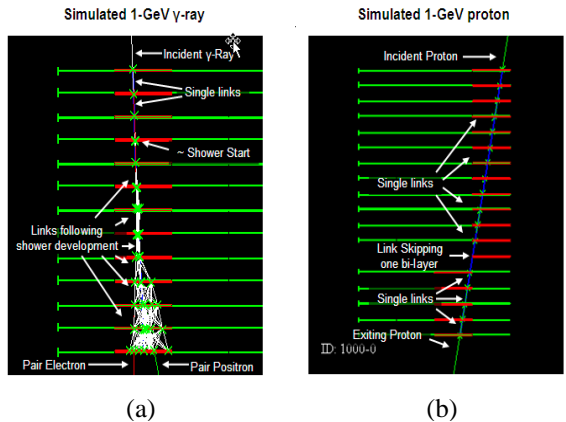

Figure 12: a) the vector links for a simulated 1-GeV gamma; b) the links for a 1-GeV proton. The EM shower in the gamma generates complex links, in contrast to the proton.

As a further step in the refinement of the track-finding, we are trying to extract the underlying "tree" from the structure of linked vectors. An example of an event with its linked vector structure and the resulting tree is shown in Fig. 13.

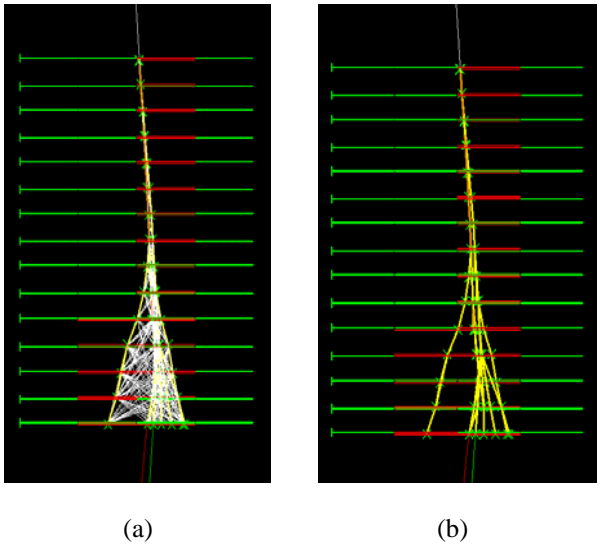

Figure 13: (a) Structure of linked vectors for a 1-GeV simulated gamma; (b) same event, showing the tree extracted from the links.

Initially, we are using this approach to find individual tracks, in order to compare this new algorithm to the current one. Even at this early stage, the results appear to be roughly comparable. But looking at Fig. 13b, it seems reasonable to contemplate deriving the kinematic information about the event directly from the tree, and eventually characterizing the "gamma-ness" of the event by examining the properties of the tree.

## Acknowledgments

The *Fermi* LAT Collaboration acknowledges generous ongoing support from a number of agencies and institutes that have supported both the development and the operation of the LAT as well as scientific data analysis. These include the National Aeronautics and Space Administration and the Department of Energy in the United States, the Commissariat à l'Energie Atomique and the Centre National de la Recherche Scientifique / Institut National de Physique Nucléaire et de Physique des Particules in France, the Agenzia Spaziale Italiana and the Istituto Nazionale di Fisica Nucleare in Italy, the Ministry of Education, Culture, Sports, Science and Technology (MEXT), High Energy Accelerator Research Organization (KEK) and Japan Aerospace Exploration Agency (JAXA) in Japan, and the K. A. Wallenberg Foundation, the Swedish Research Council and the Swedish National Space Board in Sweden.

Additional support for science analysis during the operations phase from the following agencies is also gratefully acknowledged: the Istituto Nazionale di Astrofisica in Italy and the Centre National d'Etudes Spatiales in France.